%% file: 0_main.tex
\documentclass[10pt,letterpaper]{article}
\usepackage[top=0.85in,left=2.75in,footskip=0.75in,marginparwidth=2in]{geometry}

\usepackage[utf8]{inputenc}

\usepackage{cite}

\usepackage{hyperref}
\usepackage{xcolor}
\hypersetup{
    colorlinks,
    linkcolor={red!50!black},
    citecolor={blue!50!black},
    urlcolor={blue!80!black}
}

\usepackage{ragged2e}

\usepackage[right]{lineno}

\usepackage{microtype}
\DisableLigatures[f]{encoding = *, family = * }

\raggedright
\usepackage{changepage}

\usepackage[aboveskip=1pt,labelfont=bf,labelsep=period,singlelinecheck=off]{caption}

\makeatletter
\renewcommand{\@biblabel}[1]{\quad#1.}
\makeatother

\usepackage{lastpage,fancyhdr,graphicx}
\usepackage{amsmath}
\usepackage{amssymb}
\usepackage{epstopdf}
\fancyheadoffset[L]{2.25in}
\fancyfootoffset[L]{2.25in}

\usepackage{color}

\definecolor{Gray}{gray}{.25}

\usepackage{graphicx}

\usepackage{sidecap}

\usepackage{wrapfig}
\usepackage[pscoord]{eso-pic}
\usepackage[fulladjust]{marginnote}
\reversemarginpar

\begin{document}
\vspace*{0.35in}

% title goes here:
\begin{flushleft}
{\Large
\textbf\newline{Finding Explanations of Entity Relatedness in Graphs: A Survey}
}
\newline
% authors go here:
\\
Raoul Biagioni\textsuperscript{1*},
Pierre-Yves Vandenbussche\textsuperscript{2},
V\'{i}t Nov\'{a}\v{c}ek\textsuperscript{1},
\\
\bigskip
\bf{1} Fujitsu Ireland Ltd., Co. Dublin, Ireland
\\
\bf{2} Insight Centre for Data Analytics at NUI Galway, Co. Galway, Ireland
\\
\bigskip
* raoul.biagioni@ie.fujitsu.com

\end{flushleft}

\section*{Abstract}\justifying
Analysing and explaining relationships between entities in a graph is a fundamental problem associated with many practical applications. For example, a graph of biological pathways can be used for discovering a previously unknown relationship between two proteins. Domain experts, however, may be reluctant to trust such a discovery without a detailed explanation as to why exactly the two proteins are deemed related in the graph. This paper provides an overview of the types of solutions, their associated methods and strategies, that have been proposed for finding entity relatedness explanations in graphs. The first type of solution relies on information inherent to the paths connecting the entities. This type of solution provides entity relatedness explanations in the form of a list of ranked paths. The rank of a path is measured in terms of importance, uniqueness, novelty and informativeness. The second type of solution relies on measures of node relevance. In this case, the relevance of nodes is measured w.r.t. the entities of interest, and relatedness explanations are provided in the form of a subgraph that maximises node relevance scores. This paper uses this classification of approaches to discuss and contrast some of the key concepts that guide different solutions to the problem of entity relatedness explanation in graphs.

\input{1_introduction}

\input{2_ranked_paths}

\input{3_1_optimized_subgraphs}

\input{3_2_maximal_relevance}

\input{3_3_minimum_cost}
\input{3_4_grouping}
\input{4_discussion}

%This is where your bibliography is generated. Make sure that your .bib file is actually called library.bib
\bibliography{literature.bib}

%This defines the bibliographies style. Search online for a list of available styles.
\bibliographystyle{abbrv}

\end{document}

%% file: 1_introduction.tex
% !TeX root = ../main.tex
%
\section{Introduction}\label{intro}

Graphs can conveniently represent large amounts of information as networks of relationships between objects. Structuring information as a graph allows to search for relationships between some entities of interest. However, in large graphs, information may range from trivial to meaningful, and a need thus exists for extracting only those relationships that are relevant and meaningful in a particular context. For example, a graph of biological pathways can be used for discovering a previously unknown relationship between two proteins. Domain experts, however, may be reluctant to trust such a discovery without a detailed explanation as to why exactly the two proteins are deemed related in the graph. Judgement of what constitutes a relationship, and what is meaningful and non-trivial is nonetheless a subjective matter. For this reason solutions have been proposed in the literature to find meaningful and non-trivial relationships between entities of interest in objective ways. In this paper we provide an overview of the types of solutions, their associated methods and strategies proposed for finding such relationships, henceforth referred to as \textit{entity relatedness explanations}. 

\subsection{Focus of this Paper}

The focus of this paper is to present and discuss strategies and methods that are able to select relevant relationships between entities of interest in a graph in order to make the knowledge embedded in a graph more usable e.g. for visualisation or analytical purposes. Although this paper does not claim to be exhaustive, it provides an overview of key methods that are representative solutions to this problem. We distinguish two solutions to the problem of finding entity relatedness explanations. The first type of solution relies on information inherent to the paths connecting the entities. This type of solution provides entity relatedness explanations in the form of a ranked list of \textit{relevant paths}. The rank of a path is measured in terms of importance, uniqueness, novelty and informativeness. The second type of solution relies on measures of node relevance. In this case, the relevance of a node is measured w.r.t. the entities of interest and relatedness explanations are provided in the form of an \textit{explanation subgraph} that maximises node relevance scores. This paper uses this classification of approaches to discuss and contrast some of the key concepts that guide different solutions to the problem of entity relatedness explanation in graphs.

 \subsection{Summary of Contributions}\label{contrib}

The main contributions of this survey are the following:

\begin{itemize}
    \item We identify two possible solutions to the problem of entity relatedness explanations in graphs. One focusing on the graph theoretic and, where applicable, semantic information inherent to the path connecting entities of interest; the second focusing on finding one or more optimal subgraphs by measuring the relevance of nodes w.r.t. entities of interest. 
    
    \item For each solution type, we provide a synthesis of core methods and strategies.
    
    \item Finally we discuss the assumptions underpinning the graph theoretic concepts from which solutions are derived and how these assumptions may or may not be applicable to the entity relatedness explanation problem.  
\end{itemize}

To the best of the authors’ knowledge only one similar survey of entity relatedness explanations in graphs exists \cite{alsudairyconnection2011}. To explain the relationships among entities of interest, the survey groups papers by whether solutions to finding relatedness explanations concern two entities of interest, or whether they concern two or more entities. In this survey we extend this view by framing the discussion around the strategies and rationale behind identifying and extracting relevant relationships between entities of interest. 

The requirement for this survey emanated from a specific biomedical research case. As part of this research a knowledge graph of biomedical entities had been assembled from multiple heterogeneous data sources. However, finding relevant relationships for two given entities of interest, among the myriad of paths connecting those entities was challenging and prompted the requirement for surveying the existing entity relatedness literature in search for possible solutions. Relevant research publications were selected through a process of reference exploration, starting from two seed papers \cite{ramakrishnandiscovering2005, pirroexplaining2015}. From the seed paper, related works references were repeatedly explored with the goal of finding papers that discuss directly or indirectly the entity relatedness explanation problem. The initial search returned 336 articles and 19 met the requirement for detailed review. 

\subsection{Organisation of this Paper}

This paper is organised around the two proposed types of solutions to the problem of finding entity relatedness explanations in a graph. Section~\ref{ranked_paths} introduces \textit{relevant paths extraction}, a solution that retrieves explanations in the form of a list of ranked paths. Section~\ref{optimized_subgraph} introduces \textit{relevant subgraph extraction}, a solution that retrieves entity relatedness in the form of optimised subgraphs. We propose to distinguish between \textit{maximal relevance} (Section~\ref{maxrel}), \textit{minimum cost} (Section~\ref{mincost}) and \textit{node grouping} (Section~\ref{grouping}) strategies. Finally, in Section~\ref{conclusion} we discuss and contrast some of the key concepts that guide both relevant paths and relevant subgraph extraction solutions.

%% file: 2_ranked_paths.tex
% !TeX root = ../main.tex
%

\section{Relevant Paths Extraction}\label{ranked_paths}

One possible solution to the entity relatedness explanation problem is to consider all paths that connect two entities of interest and to identify those paths that best explain the relationship between those entities. Relatedness of nodes in paths w.r.t. entities of interest is assumed and the task is to rank paths in terms of their relevance to the context, their uniqueness, novelty and informativeness. This is achieved by exploiting graph theoretic, statistical and semantic information embedded in paths. Relevant paths extraction solutions are limited to two entities of interest.

The research areas covered by relevant paths extraction solutions include literature-based discovery (LBD)~\cite{wilkowskigraphbased2011}, link discovery~\cite{linchalupsky2003} and the Semantic Web~\cite{alemanmezaranking2005, anyanwusemrank2005, pirroexplaining2015, ramakrishnandiscovering2005}. Examples of input graphs used include co-citation graphs~\cite{linchalupsky2003} and biochemical networks~\cite{wilkowskigraphbased2011}. Input graphs used in Semantic Web research papers are based on the Resource Description Framework (RDF), a data model based on named relationships between resources~\cite{alemanmezaranking2005}. A well known example of an RDF-based input graph is \textit{Freebase} and \textit{DBPedia}. 

The primary graph theoretic measures used for path ranking are node degree and path length. Node degree is considered indicative of e.g. activity in a social network or interaction in a genetic regulatory network. High node degree suggests importance and paths containing such nodes are ranked higher. For example, in \cite{wilkowskigraphbased2011}, 
the importance of a given path is scored using the arithmetic sum of the degree centrality values $dc$ for all $n$ nodes $B$ in the path (Eq. \ref{sumnodedegrees}).

\begin{equation}\label{sumnodedegrees}
score = \sum_{n} dc(B_n)
\end{equation}

The length of paths connecting entities of interest is used as a proxy measure for association strength. The shorter a path, the stronger the association. However, a strong association does not necessarily lead to a higher rank. In certain contexts one may want to give preference to weak associations i.e. long paths. For example, as described in~\cite{alemanmezaranking2005}, money laundering involves transactions that may change several hands. Hence, in a graph representing financial transactions, one may want to include long paths in a relatedness explanation for sender and receiver entities. For this reason, the authors define the weight $L$ of a path $A$ such that the contribution of a path's length to its rank be informed by whether preference should be given to short (Eq. \ref{associationstrength_a}) or long paths (Eq. \ref{associationstrength_b}), where $length(A)$ is the number of nodes in A.

\begin{subequations}  
\begin{align}   
L_A = \frac{1}{length(A)} \label{associationstrength_a}\\   
L_A = 1 - \frac{1}{length(A)} \label{associationstrength_b} 
\end{align}  
\end{subequations}

Statistical path scoring measures are based on frequency of occurrence counts, e.g. the number of similar paths~\cite{linchalupsky2003}, the number of nodes in a path~\cite{alemanmezaranking2005}, or the number of edges in a path~\cite{anyanwusemrank2005}. In~\cite{linchalupsky2003} and~\cite{anyanwusemrank2005}, low frequency of occurrence counts represent \textit{rarity} and \textit{specificity}. Similarly, in~\cite{pirroexplaining2015}, high frequency of occurrence counts represent \textit{diversity}. In this case, the authors use the frequency of occurrence of nodes and edges to define \textit{path diversity} and \textit{path informativeness} measures. Path informativeness is based on the frequency of occurrence of edges and the inverse frequency of occurrence of nodes~\cite{pirroexplaining2015}. Path diversity $\sigma$ is measured by the ratio of common labels to the total number of labels between two paths $\pi_1$ and $\pi_2$ (Eq. \ref{labeldiversity}). Both measures are complementary. Path diversity ensures that paths with rare edges, potentially discarded if they appear in paths with low informativeness, are retained. As illustrated by this example, statistics-based ranking is typically framed in terms of conceptual ideas. The properties of information associated with those concepts are novelty, uniqueness, non-triviality and informativeness. In general, those properties are deemed desirable and result in a higher rank. 

\begin{equation}\label{labeldiversity}
\sigma(\pi_1, \pi_2)=\frac{\left | Labels(\pi_1) \cap Labels(\pi_2) \right |}{\left | Labels(\pi_1) \cup  Labels(\pi_2) \right |}
\end{equation}

Path ranking measures are often used in combination. For example, in \cite{alemanmezaranking2005}, the overall frequency of occurrence $rar_i$ of a node $i$ in an input graph is combined with the association length weight $L_A$ described above to produce an overall weight of path rarity $R_A$ (Eq.~\ref{rarityweight_a} for short and Eq.~\ref{rarityweight_b} for long paths).

\begin{subequations}  
\begin{align}   
R_A = \frac{1}{length(A)} \sum_{i=1}^{length(A)} rar_i \label{rarityweight_a}\\   
R_A = 1 - \frac{1}{length(A)} \sum_{i=1}^{length(A)} rar_i \label{rarityweight_b} 
\end{align}  
\end{subequations}

Similarly, \cite{anyanwusemrank2005} and \cite{ramakrishnandiscovering2005}  combine measures of \textit{specificity} and of \textit{diversity}. Specifically, where a graph is composed of multiple schemas (i.e. domains), this information can be used to determine whether a given path ``spans'' several domains. The more schemas are covered by the edges in a path, the higher the rank. Lastly, semantic information is extracted, where applicable, from RDF~\cite{ramakrishnandiscovering2005} and ontology schema hierarchy~\cite{alemanmezaranking2005}. Such hierarchies represent a partial ordering of entities and, the deeper an entity lies in the hierarchy, the more specific or informative it is.

%% file: 3_1_optimized_subgraphs.tex
% !TeX root = ../main.tex
%

\section{Relevant Subgraph Extraction}\label{optimized_subgraph}

A second approach to solving the entity relatedness explanation problem is to extract a subgraph such that its nodes are as relevant as possible to some entities of interest. We distinguish between strategies that focus on \textit{maximal relevance} (Section \ref{maxrel}), \textit{minimum cost} (Section \ref{mincost}) and \textit{node grouping} (Section \ref{grouping}) methods. Relevant subgraph extraction is suited to cases with more than two entities of interest e.g. $Q=\{q_1,...,q_n\}$ or with two sets of entities $Q1$ and $Q2$. 

The research areas covered by 
%papers that propose 
relevant subgraph extraction solutions include data mining~\cite{liupeak2016}, knowledge discovery~\cite{kasneciming2009}, graph mining~\cite{ruchansky2017}, graph theory~\cite{akoglumining2013} and graph databases~\cite{chengcorrelationgroups2009, chengcontextaware2009}. Examples of input graphs used in the reviewed papers include domain-specific knowledge bases, social networks, food networks, protein-protein interaction networks, co-authorship graphs.

%% file: 3_2_maximal_relevance.tex
% !TeX root = ../main.tex
%

\subsection{Maximal Relevance Approaches}\label{maxrel}

Maximal relevance approaches select an explanation subgraph by maximising a \textit{goodness function} that measures and optimises the relevance of \textit{candidate subgraphs}. The relevance of a subgraph is assessed through the relevance and/or importance of its constituent nodes. Relevance relates to the entities of interest, importance relates to the network as a whole. A globally important node may be one that acts as an important information bridge, authority or hub;  as opposed to nodes that are isolated, with little influence on the rest of the network. The importance and relevance of nodes is used to search for the most informative explanation subgraph. The computation of importance and relevance scores is discussed next.

\subsubsection{Node Scoring}\label{nodescor}

Node relevance scores are typically computed using \textit{random walk} (RW)~\cite{chensisp2011, chengcorrelationgroups2009, dupontrelevant2006, kasneciming2009, seufertespresso2016, tongcenterpiece2006,wangpathprobability2013}. The RW method allows to compute a probability distribution over the set of nodes in a graph. This probability information can be used to measure node relevance i.e. how much each node contributes to the relationships with the nodes of interest. RW-based methods are also suitable for measuring the overall, or global, importance of nodes~\cite{chensisp2011, liupeak2016} in a graph. In this case a small probability is assigned to the walker ``jumping'' to any node in the graph, rather than moving to an adjacent node. A well-known example is \textit{PageRank}~\cite{pagepagerank1999}. 

The most commonly used random walk ``flavour'' is \textit{Random Walk with Restart} (RWR)
~\cite{chensisp2011, chengcorrelationgroups2009, kasneciming2009, liupeak2016, wangpathprobability2013}. Another similar type of random walk, the \textit{Absorbing Random Walk} (ARW) is proposed in~\cite{dupontrelevant2006}. In a RWR, a small probability is assigned to the walker restarting the walk from where it began. This restart probability makes the RWR method an attractive choice in light of the ``w.r.t. entities of interest'' aspect of node relevance scoring. In an ARW, a walker stops (i.e. is absorbed) when an entity of interest is reached. ARW is used to compute the expected number of times a node is used when randomly walking through the graph, starting from one entity of interest until eventually being absorbed by a distinct entity of interest. The relevance of a node is proportional to these quantities. 

The relevance $r$ of a node $j$ w.r.t. the entire set $Q$ of entities of interest $q_{i} \in Q$ can be computed, as seen in e.g. \cite{tongcenterpiece2006}, by multiplying individual pair-wise relevance scores (Eq.~\ref{relproduct}). 

\begin{equation}\label{relproduct}
r(Q,j)= \prod_{i=1}^{Q}{r(i,j)} 
\end{equation}\textit{}

The above approach has, however, one potential weakness. It assumes that candidate explanatory nodes have a relationship with \textit{all} entities of interest. This can be addressed by relaxing the way in which node relevance w.r.t. the entire set of entities of interest is computed. For example, ~\cite{tongcenterpiece2006} allow for two additional scenarios. Node relevance may relate to \textit{at least one} (Eq.~\ref{atleastone}) or to \textit{at least k} entities of interest, where $k~(1 \leq k \leq Q)$. The same idea is adopted in~\cite{seufertespresso2016}. In their paper, the authors propose entity relatedness explanations for two sets of entities of interest $Q1$ and $Q2$. The authors search for ``central nodes'' to which entities of interest are related in some relevant way e.g. through common geo-political events. An essential aspect of the methodology consists of relaxing the stringency requirements on how many of the entities of interest in each of the sets must exhibit such commonalities with the central nodes. 

\begin{equation}\label{atleastone}
r(Q,j) \triangleq r(Q,j,1)= 1- \prod_{i=1}^{Q}{(1-r(i,j))}
\end{equation}

\subsubsection{Subgraph Optimisation}\label{goodness}

Methods that provide entity relatedness explanations in the form of a subgraph \textit{H}~\cite{chensisp2011, faloutsosfast2004, liupeak2016, tongcenterpiece2006} are designed such as to maximise the ``goodness" of this subgraph (Eq.~\ref{goodnessfunc}). The goodness of $H$ is measured by the sum of the relevance scores of the nodes it contains (Eq.~\ref{goodnessfunc2}) where $r(Q,j)$ is the relevance score of node $j$ w.r.t. all nodes of interest in $Q$.

\begin{equation}\label{goodnessfunc}
H^*=argmax_{H}g(H)
\end{equation}

\begin{equation}\label{goodnessfunc2}
g(H) = \sum_{j \in H}{r(Q,j)}
\end{equation}

The above approach can easily be adapted to cases where the relatedness explanation is composed of more than one subgraph. This case is illustrated in \cite{liupeak2016}. The authors partition the input graph into $k$ groups ($P_1\ to\ P_k$). In this case, the goodness of $H$ is measured by the sum of the maximised goodness score of each group $P_i$ where $Q_{P_{i}}'$ is a subset of $Q$ and all the nodes in $Q_{P_{i}}'$ belong to $P_{i}$ (Eq. \ref{group_goodness}). 

\begin{equation}\label{group_goodness}
g(H) = \sum_{P_{i}} \sum_{j \in P_{i}}{r(j,Q_{P_{i}}')}
\end{equation}

%% file: 3_3_minimum_cost.tex
% !TeX root = ../main.tex
%

\subsection{Minimum Cost Approaches}\label{mincost}

Minimum cost approaches select relevant subgraphs based on the assumption that related nodes are close to each other. The proposition made is that entity relatedness explanations can be extracted by interconnecting the entities of interest through shortest paths, subject to the input graph being fully connected. The minimum cost approach minimises the number of edges (for unweighted graphs) or the weights of the edges (for weighted graphs) in the subgraph. The resulting subgraph is known as a \textit{spanning tree} and the problem of producing a spanning tree for a set of entities of interest is known as the \textit{Steiner Tree problem}. Authors that frame the entity relatedness explanation problem as a Steiner Tree problem include \cite{akoglumining2013, chengcorrelationgroups2009, kasneciming2009, ruchansky2017}. 

%% file: 3_4_grouping.tex
% !TeX root = ../main.tex
%

\subsection{Node Grouping}\label{grouping}

As described in Section \ref{nodescor}, explanatory nodes may not necessarily be related to \textit{all} entities of interest and a possible solution for dealing with this scenario is to relax relevance requirements. In this section, we briefly describe an alternative solution, namely grouping nodes, that have high relevance w.r.t. each other~\cite{akoglumining2013, chengcorrelationgroups2009, chengcontextaware2009,liupeak2016}. In this case, the intention is to spread entities of interest across groups. This then allows for connecting paths to be restricted to a subset of entities of interest at a time before inter-connecting the groups. A diverse and eclectic set of methods has been proposed in the literature to achieve this task. In~\cite{chengcorrelationgroups2009}, nodes are grouped by their correlation to one of the entities of interest. For each entity of interest $q$, the authors find a set of nodes that are pair-wise correlated to $q$. A pair of nodes $u$ and $v$ are correlated if $r(u,v) \geq \sigma$ and $r(v,u) \geq \sigma$, where $\sigma$ is a predefined minimum relevance threshold. Other methods include clustering nodes into domain-specific communities~\cite{chengcontextaware2009}, 
grouping nodes based on density of linkages~\cite{liupeak2016}, 
methods based on Information Theory~\cite{akoglumining2013} and methods based on flow information in networks~\cite{chengcontextaware2009}. Approaches to inter-connecting groups of related nodes include minimum cost methods~\cite{chengcorrelationgroups2009}.

%% file: 4_discussion.tex
% !TeX root = ../main.tex
%

\section{Discussion and Conclusion}\label{conclusion}

In this paper, we presented a survey of strategies and methods for finding entity relatedness explanations in graphs. We identified two possible solutions to this problem. One focusing on graph theoretic and semantic information embedded in paths connecting entities of interest; the second focusing on finding an optimal subgraph in which the relevance scores of nodes w.r.t. entities of interest is maximised. In this Section, we discuss and contrast some of the key concepts that guide both the relevant paths and relevant subgraph extraction solutions.

\subsection{Relevance and Shortest Paths}\label{conclusion1}

An interesting observation made in the process of this survey is the lack of adoption of some of the traditional node centrality measures such as 
closeness~\footnote{\textit{Closeness} is the mean shortest-path distance between a node and all other nodes reachable from it~\cite{newman2003}.} 
and 
betweenness~\footnote{Betweenness is calculated as the fraction of shortest paths between node pairs that pass through the node of interest~\cite{newman2003}.}. 
This may appear surprising at first considering how node centrality is a fundamental measure for understanding the roles played by nodes in networks~\cite{newman2003}. The lack of adoption can be explained by the fact that graph centrality measures assume that information travels along shortest paths. However, adopting this assumption without considering the nature of the information being represented as a graph may be problematic. For example, in an industry-news network on the one hand, the shortest path is known and information travels in a targeted manner. On the other hand, in a social network, gossip propagates without specific target; using shortest path-based centrality measures to explain the flow of gossip is likely to produce low quality results~\cite{borgatticentrality2005}. Similarly, in many of the domains in which entity relatedness explanation are sought, information does not take any sort of ideal path from source to target and, in such cases, node centrality measures could fail to adequately model entity relationships~\cite{faloutsosfast2004}.
 
Whether shortest paths methods are deemed applicable to the entity relatedness problem differs among authors. For example, in~\cite{chensisp2011} and in~\cite{kasneciming2009}, shortest paths are used as indicators of relevance that adequately capture entity relatedness while in~\cite{anyanwusemrank2005} preference is given to long paths because they are seen as more specific thus ``more likely to reveal rare and uncommon associations'' between entities of interest. The solution proposed in \cite{alemanmezaranking2005} allows for both views. Shorter paths are seen as representing direct relationships, while longer paths represent indirect relationships. Both direct and indirect relationships can be meaningful, depending on the problem at hand, and it is up to the practitioners to define their preference. 

How can the ambivalence concerning the use of shortest paths as means of extracting relatedness explanations from graphs be addressed? Is it more useful to try extracting relatedness explanations by modelling information flow as a random process? Indeed, in~\cite{dupontrelevant2006} and in~\cite{newman2003} it is argued that, if information flows randomly, then contributions from all paths, not just the shortest, should contribute to entity relatedness explanations. It is precisely this observation which led to the prevalence of RW-based methods in relevant subgraph extraction methods. RW-based methods allow to compute node relevance scores on all possible ways to connect entities of interest, each way having a certain likelihood, rather than relying only on shortest distance.

\subsection{The Subjective Nature of Relevance}

Whether a relatedness explanation is relevant is, as noted in Section~\ref{intro}, a matter of subjective judgement. Different people will almost certainly have differing notions of relevance. An entity relatedness explanation grounded in LBD will need to be based on the uncommon and novel in order to support the discovery of new, yet unknown relationships. In this case, solutions and methods that favour nodes in a graph that have low degree will be preferred because high degree can be seen as a weak measure because a node through which many entities are connected is too general for contributing to succinctly explain how entities of interest are related~\cite{shiralkarfinding2017}. Conversely, an entity relatedness explanation which deals with biological knowledge such as a gene interaction graph will need to take into account that high connectivity is an important property of the underlying graph~\cite{wilkowskigraphbased2011}. In this case, high node degree is desirable and is seen as an appropriate indicator of relevance. 

The above example illustrates that there is no one correct answer to the question whether high, or low, node degree is a good indicator of relevance. The advantage of relevant paths extraction over relevant subgraph extraction lies in that the former allows to more easily adapt to the configurations deemed most appropriate for a given context. Specifically, entity relatedness explanations are configurable through the path ranking scheme inherent to the relevant paths extraction. Examples of papers that implemented relevant paths extraction solutions as configurable, user-driven and interactive applications include~\cite{alemanmezaranking2005, pirroexplaining2015, ramakrishnandiscovering2005, anyanwusemrank2005}. In contrast, relevant subgraph extraction allows for only limited ``tuning'' to account for subjective notions of relevance. The advantages of this type of solution over relevant paths extraction are two-fold. First, relevant subgraph extraction methods allow for more than two entities of interest to be specified. Second, as discussed in Section \ref{conclusion1}, the way in which information flow is modelled, namely as a random process, can be seen as a more adequate way of representing knowledge in graphs.

\subsection{Final Remarks}

In Section~\ref{contrib}, we described how the requirement for this survey arose from the challenges associated with manually inspecting the myriad of connecting paths between entities of interest in a graph. The solutions, methods and strategies described in this paper can help practitioners make decisions in a more informed way when faced with implementing solutions to the task of explaining the relationships between entities of interest. Based on the input graph at hand, the particularities of the context and the information represented, our paper helps practitioners assess which one of the two types of solutions, and their corresponding methods and strategies best suits their needs and context.  

\section*{Acknowledgements}

The work presented in this paper was supported by the TOMOE project funded by Fujitsu Laboratories Ltd., Japan and Insight Centre for Data Analytics at National University of Ireland Galway (supported by the Science Foundation Ireland grant 12/RC/2289).

%% file: 0_main.bbl
\begin{thebibliography}{10}

\bibitem{akoglumining2013}
L.~Akoglu, D.~H. Chau, J.~Vreeken, N.~Tatti, H.~Tong, and C.~Faloutsos.
\newblock Mining {Connection} {Pathways} for {Marked} {Nodes} in {Large}
  {Graphs}.
\newblock In {\em Proceedings of the 2013 {SIAM} {International} {Conference}
  on {Data} {Mining}}, pages 37--45. Society for Industrial and Applied
  Mathematics, 2013.

\bibitem{alemanmezaranking2005}
B.~Aleman-Meza, C.~Halaschek-Weiner, I.~B. Arpinar, C.~Ramakrishnan, and A.~P.
  Sheth.
\newblock Ranking complex relationships on the semantic {Web}.
\newblock {\em IEEE Internet Computing}, 9(3):37--44, 2005.

\bibitem{alsudairyconnection2011}
N.~M.~K. Alsudairy, V.~V. Raghavan, A.~M. Hafez, and H.~I. Mathkour.
\newblock Connection {Subgraphs}: {A} {Survey}.
\newblock {\em Journal of Applied Sciences}, 11:3221--3232, 2011.

\bibitem{anyanwusemrank2005}
K.~Anyanwu, A.~Maduko, and A.~Sheth.
\newblock {SemRank}: {Ranking} {Complex} {Relationship} {Search} {Results} on
  the {Semantic} {Web}.
\newblock In {\em Proceedings of the 14th {International} {Conference} on
  {World} {Wide} {Web}}, {WWW} '05, pages 117--127. ACM, 2005.

\bibitem{borgatticentrality2005}
S.~P. Borgatti.
\newblock Centrality and network flow.
\newblock {\em Social Networks}, 27(1):55--71, 2005.

\bibitem{chensisp2011}
C.~Chen, G.~Wang, H.~Liu, J.~Xin, and Y.~Yuan.
\newblock {SISP}: {A} {New} {Framework} for {Searching} the {Informative}
  {Subgraph} {Based} on {PSO}.
\newblock In {\em Proceedings of the 20th {ACM} {International} {Conference} on
  {Information} and {Knowledge} {Management}}, {CIKM} '11, pages 453--462. ACM,
  2011.

\bibitem{chengcorrelationgroups2009}
J.~Cheng, Y.~Ke, and W.~Ng.
\newblock Efficient {Processing} of {Group}-oriented {Connection} {Queries} in
  a {Large} {Graph}.
\newblock In {\em Proceedings of the 18th {ACM} {Conference} on {Information}
  and {Knowledge} {Management}}, {CIKM} '09, pages 1481--1484. ACM, 2009.

\bibitem{chengcontextaware2009}
J.~Cheng, Y.~Ke, W.~Ng, and J.~X. Yu.
\newblock Context-{Aware} {Object} {Connection} {Discovery} in {Large}
  {Graphs}.
\newblock In {\em 2009 {IEEE} 25th {International} {Conference} on {Data}
  {Engineering}}, pages 856--867, 2009.

\bibitem{dupontrelevant2006}
P.~Dupont, J.~Callut, G.~Dooms, J.-N. Monette, Y.~Deville, and B.~P. Sainte.
\newblock Relevant subgraph extraction from random walks in a graph.
\newblock {\em Universite Catholique de Louvain, UCL/INGI, Research Report RR},
  380167, 2006.

\bibitem{faloutsosfast2004}
C.~Faloutsos, K.~S. McCurley, and A.~Tomkins.
\newblock Fast discovery of connection subgraphs.
\newblock In {\em Proceedings of the Tenth ACM SIGKDD International Conference
  on Knowledge Discovery and Data Mining}, KDD '04, pages 118--127. ACM, 2004.

\bibitem{kasneciming2009}
G.~Kasneci, S.~Elbassuoni, and G.~Weikum.
\newblock {MING}: {Mining} {Informative} {Entity} {Relationship} {Subgraphs}.
\newblock In {\em Proceedings of the 18th {ACM} {Conference} on {Information}
  and {Knowledge} {Management}}, {CIKM} '09, pages 1653--1656. ACM, 2009.

\bibitem{linchalupsky2003}
S.-d. Lin and H.~Chalupsky.
\newblock Unsupervised link discovery in multi-relational data via rarity
  analysis.
\newblock In {\em Third {IEEE} {International} {Conference} on {Data}
  {Mining}}, pages 171--178, 2003.

\bibitem{liupeak2016}
H.~Liu, C.~Chen, J.~Xin, and L.~Zhang.
\newblock Searching the {Informative} {Subgraph} {Based} on the {PeakGraph}
  {Model}.
\newblock {\em The Computer Journal}, 59(8):1207--1219, 2016.

\bibitem{newman2003}
M.~E.~J. Newman.
\newblock A measure of betweenness centrality based on random walks.
\newblock {\em Social Networks}, 27(1):39 -- 54, 2005.

\bibitem{pagepagerank1999}
L.~Page, S.~Brin, R.~Motwani, and T.~Winograd.
\newblock The {PageRank} citation ranking: {Bringing} order to the web.
\newblock Technical report, Stanford InfoLab, 1999.

\bibitem{pirroexplaining2015}
G.~Pirr\'o.
\newblock Explaining and {Suggesting} {Relatedness} in {Knowledge} {Graphs}.
\newblock In {\em The {Semantic} {Web} - {ISWC} 2015}, Lecture {Notes} in
  {Computer} {Science}, pages 622--639. Springer, Cham, 2015.

\bibitem{ramakrishnandiscovering2005}
C.~Ramakrishnan, W.~H. Milnor, M.~Perry, and A.~P. Sheth.
\newblock Discovering {Informative} {Connection} {Subgraphs} in
  {Multi}-relational {Graphs}.
\newblock {\em ACM SIGKDD Explorations Newsletter}, 7(2):56--63, 2005.

\bibitem{ruchansky2017}
N.~Ruchansky, F.~Bonchi, D.~Garcia-Soriano, F.~Gullo, and N.~Kourtellis.
\newblock To be connected, or not to be connected: That is the minimum
  inefficiency subgraph problem.
\newblock In {\em Proceedings of the 2017 ACM on Conference on Information and
  Knowledge Management}, CIKM '17, pages 879--888. ACM, 2017.

\bibitem{seufertespresso2016}
S.~Seufert, K.~Berberich, S.~J. Bedathur, S.~K. Kondreddi, P.~Ernst, and
  G.~Weikum.
\newblock {ESPRESSO}: {Explaining} {Relationships} {Between} {Entity} {Sets}.
\newblock In {\em Proceedings of the 25th {ACM} {International} on {Conference}
  on {Information} and {Knowledge} {Management}}, {CIKM} '16, pages 1311--1320.
  ACM, 2016.

\bibitem{shiralkarfinding2017}
P.~Shiralkar, A.~Flammini, F.~Menczer, and G.~L. Ciampaglia.
\newblock Finding streams in knowledge graphs to support fact checking.
\newblock In {\em 2017 {IEEE} International Conference on Data Mining, {ICDM}
  2017, New Orleans, LA, USA, November 18-21, 2017}, pages 859--864, 2017.

\bibitem{tongcenterpiece2006}
H.~Tong and C.~Faloutsos.
\newblock Center-piece {Subgraphs}: {Problem} {Definition} and {Fast}
  {Solutions}.
\newblock In {\em Proceedings of the 12th {ACM} {SIGKDD} {International}
  {Conference} on {Knowledge} {Discovery} and {Data} {Mining}}, {KDD} '06,
  pages 404--413. ACM, 2006.

\bibitem{wangpathprobability2013}
H.~Wang, C.~K. Chang, H.-I. Yang, and Y.~Chen.
\newblock Estimating the {Relative} {Importance} of {Nodes} in {Social}
  {Networks}.
\newblock {\em Journal of Information Processing}, 21(3):414--422, 2013.

\bibitem{wilkowskigraphbased2011}
B.~Wilkowski, M.~Fiszman, C.~M. Miller, D.~Hristovski, S.~Arabandi,
  G.~Rosemblat, and T.~C. Rindflesch.
\newblock Graph-{Based} {Methods} for {Discovery} {Browsing} with {Semantic}
  {Predications}.
\newblock {\em AMIA Annual Symposium Proceedings}, 2011:1514--1523, 2011.

\end{thebibliography}
